\documentclass[%
 aip,
 amsmath,amssymb,
 reprint,%
]{revtex4-1}

\usepackage{graphicx}% Include figure files
\usepackage{dcolumn}% Align table columns on decimal point
\usepackage{bm}% bold math
\usepackage[utf8]{inputenc}
\usepackage[T1]{fontenc}
\usepackage{mathptmx}

\begin{document}

%\preprint{AIP/123-QED}

\title[Learning the energy curvature versus particle number in approximate density functionals]{Learning the energy curvature versus particle number in approximate density functionals}
% Force line breaks with \\

\author{Alberto Fabrizio}
\author{Benjamin Meyer}%
\author{Clemence Corminboeuf}
 \email{clemence.corminboeuf@epfl.ch}
 \affiliation{ Laboratory for Computational Molecular Design, Institute of Chemical Sciences and Engineering, \'{E}cole Polytechnique F\'{e}d\'{e}rale de Lausanne, CH-1015 Lausanne, Switzerland 
}%
\affiliation{ National Centre for Computational Design and Discovery of Novel Materials (MARVEL), \'{E}cole Polytechnique F\'{e}d\'{e}rale de Lausanne, 1015 Lausanne, Switzerland 
}%

\date{\today}% It is always \today, today,

\begin{abstract}
The average energy curvature in function of the particle number is a molecule-specific quantity, which measures the deviation of a given functional from the exact conditions of density functional theory (DFT). Related to the lack of derivative discontinuity in approximate exchange-correlation potentials, the information about the curvature has been successfully used to restore the physical meaning of Kohn-Sham orbital eigenvalues and to develop non-empirical tuning and correction schemes for density functional approximations. In this work, we propose the construction of a machine-learning framework targeting the average energy curvature between the neutral and the radical cation state of thousands of small organic molecules (QM7 database). The applicability of the model is demonstrated in the context of system specific gamma-tuning of the LC-$\omega$PBE functional and validated against the molecular first ionization potentials at equation-of-motion (EOM) coupled-cluster references. In addition, we propose a local version of the non-linear regression model and demonstrate its transferability and predictive power by determining the optimal range-separation parameter for two large molecules relevant for the field of hole-transporting materials. Finally, we explore the underlying structure of the QM7 database with the t-SNE dimensionality-reduction algorithm and identify structural and compositional patterns that promote the deviation from the piecewise linearity condition.
\end{abstract}

\maketitle

\section{\label{sec:level1}Introduction}

The extension of Hohenberg-Kohn density functional theory (HK-DFT)\cite{Hohenberg1964} to non-integer particle numbers led to the determination of two fundamental properties of exact DFT.\cite{Perdew1982} The first is the piecewise linearity condition, which imposes that the total energy in function of the (fractional) particle number [\textit{E(N)}] must evolve as a series of straight-line segments.\cite{Perdew1982,Cohen2008,Stein2012,Kraisler2013} The second is the derivative discontinuity, which establishes that the exact exchange-correlation potential is characterized by sudden jumps while varying across integer particle numbers.\cite{Perdew1982,Mundt2005,Mori-Sanchez2009ab,Mirtschink2013,Yang2012,Mori-Sanchez2014}

Approximate density functionals do not fulfill these requirements. Instead, they are generally characterized by a convex \textit{E(N)} curvature and by continuously derivable exchange-correlation potentials.\cite{Mori-Sanchez_J.Chem.Phys.2006,Ruzsinszky2007,Vydrov2007,Mori-Sanchez2008,Cohen2008a,Cohen2008,Haunschild2010} As demonstrated by Kronik, Baer and coworkers,\cite{Stein2012} these two quantities are related and therefore the knowledge of the first is sufficient to quantify the extent of the second. Using the same argument, the minimization of the energy curvature has the consequence of correcting the effects of the missing derivative discontinuity, restoring the compliance of approximate functionals to the exact conditions of DFT. Failure to comply with these requirements exacerbates the effects of the delocalization error,\cite{Tozer2003,Mori-Sanchez2008,Cohen2008,Cohen_Chem.Rev.2012,Zhao2016} leads to an incorrect dissociation behavior of heterodimers\cite{Perdew1982,Zhang1998,Ruzsinszky2007} and causes the Kohn-Sham frontier orbital eigenvalues to deviate respectively from the ionization potential and the electronic affinity.\cite{Perdew1983,ALLEN2002,Kummel2008,Gorling2015}

The existence of a relationship between the curvature and the derivative discontinuity is especially convenient, as the first can be readily evaluated for a given functional and chemical system according to the following expression:\cite{Stein2012}

\begin{equation}
\label{eq:curve}
    C^N_{avg}=\int_{N-1}^{N}C^N(x) dx = \epsilon_{HOMO}^{N} - \epsilon_{LUMO}^{N-1},
\end{equation}

where $C^N_{avg}$ represents the average curvature between two integer point with $N$ and $N-1$ electrons, while $\epsilon_{HOMO}^{N}$ and $\epsilon_{LUMO}^{N-1}$ are the eigenvalues of the frontier orbitals of the $N$ and $N-1$ particle states. Equation \ref{eq:curve} is exact and it is a direct consequence of the Janak's theorem.\cite{Janak1978,Perdew1983}

Straightforward accessibility to information about the energy curvature, as well as its relation to some fundamental pitfalls of approximate density functionals, has already proven to be fertile ground and is used in numerous applications. For instance, the minimization of $C^N_{avg}$ serves as a formally motivated criterion for the compound-specific optimal tuning of range-separated hybrid density functionals.\cite{Baer2010,Gledhill2013} The accuracy of such functionals has been largely demonstrated in the computations of outer-valence spectra,\cite{Egger2014} optical rotations,\cite{Srebro2012} fundamental and optical gaps.\cite{Refaely-Abramson2012,Kronik2012,Sun2014} The energy curvature has been also applied as a criterion to assess the extent of delocalization error in approximate functionals and to rationalize, on this basis, their relative accuracy.\cite{Johnson2013,Autschbach2014,Whittleton2015,Ioannidis2015,Gani2016,Hait2018} Within a different context, knowledge of the curvature played a central role in the validation of ensemble generalizations of standard density functionals,\cite{Kraisler2013,Kraisler2014,Kraisler2015,Kraisler2015a} carefully designed to retrieve the correct piecewise-linearity behavior of \textit{E(N)} and the derivative discontinuities in the exchange-correlation potential. Finally, information on the curvature is exploited to develop correction schemes for existing exchange-correlation density functionals.\cite{Cococcioni2005,Dabo2010,Stein2012,Chai2013,Ferretti2014} 

The relevance of the information encoded in the energy curvature, corroborated by the extent of possible applications, contrasts with the modest chemical complexity and relatively low number of molecules for which the curvature has been reported.\cite{Stein2012,Whittleton2015} 

Recently, machine-learning (ML) techniques have been redefining the scale and the complexity of achievable quantum chemical tasks.\cite{Blum2009,Rupp2012,Ramakrishnan2014,Faber2016,Faber2017,lilienfeld2019exploring,Liang2019,Butler2018,Fabrizio2019Chimia} Supported by the construction of large molecular databases of quantum chemical benchmarks,\cite{Blum2009,Rupp2012,Ruddigkeit2012,Montavon2013,Ramakrishnan2014,McGibbon2017,Burns2017} supervised learning approaches promote the large-scale screening of any targeted molecular quantity ranging from simple ground-state properties\cite{VonLilienfeld2018} to complex objects such as electron densities\cite{Brockherde2017,Grisafi2019,Chandrasekaran2019,Fabrizio-CHEMSCI-2019} and the many-body-wavefunction.\cite{schtt2019unifying} Tackling the scaling up through with artificial intelligence techniques is especially appealing. Specifically, AI results in a reduction in the computational cost of accessing molecular properties,\cite{Rupp2015,VonLilienfeld2018,Ramakrishnan2017} facilitates the extrapolation of acquired information to larger and more complex chemical systems,\cite{Rupp2015a,Bartok2017} and promote the analysis and identification of non-trivial similarity patterns in otherwise unimaginably large amounts of data.\cite{Ceriotti2019}

Within this context, we here report the construction of a machine-learning model of the average energy curvature ($C^N_{avg}$) of a set of 7165 organic molecules taken from the QM7 database.\cite{Blum2009,Rupp2012} In this work, the focus is placed on the curvature between the neutral and the first radical cation state of each molecule, as its minimization leads to compliance with the Koopmans' theorem.\cite{Baer2010} The applicability of the regression framework is demonstrated by performing system-specific optimal tuning of the LC-$\omega$PBE functional\cite{Vydrov2006,Vydrov2006a,Vydrov2007} on the basis of the predicted curvatures. In addition, the transferability of the model is tested by predicting the optimal range-separation parameters and estimating the ionization potential of two larger molecules of practical use, relevant for the field of hole transport materials. Finally, we address the question whether specific chemical patterns are more prone to deviation from piecewise linearity using unsupervised dimensionality reduction algorithms to draw statistically robust relationships between the structure/composition of the molecules and their average energy curvature.

\section{\label{sec:level2}Learning Curves}

\begin{figure}[!t]
    \centering
    \includegraphics[width=0.45\textwidth]{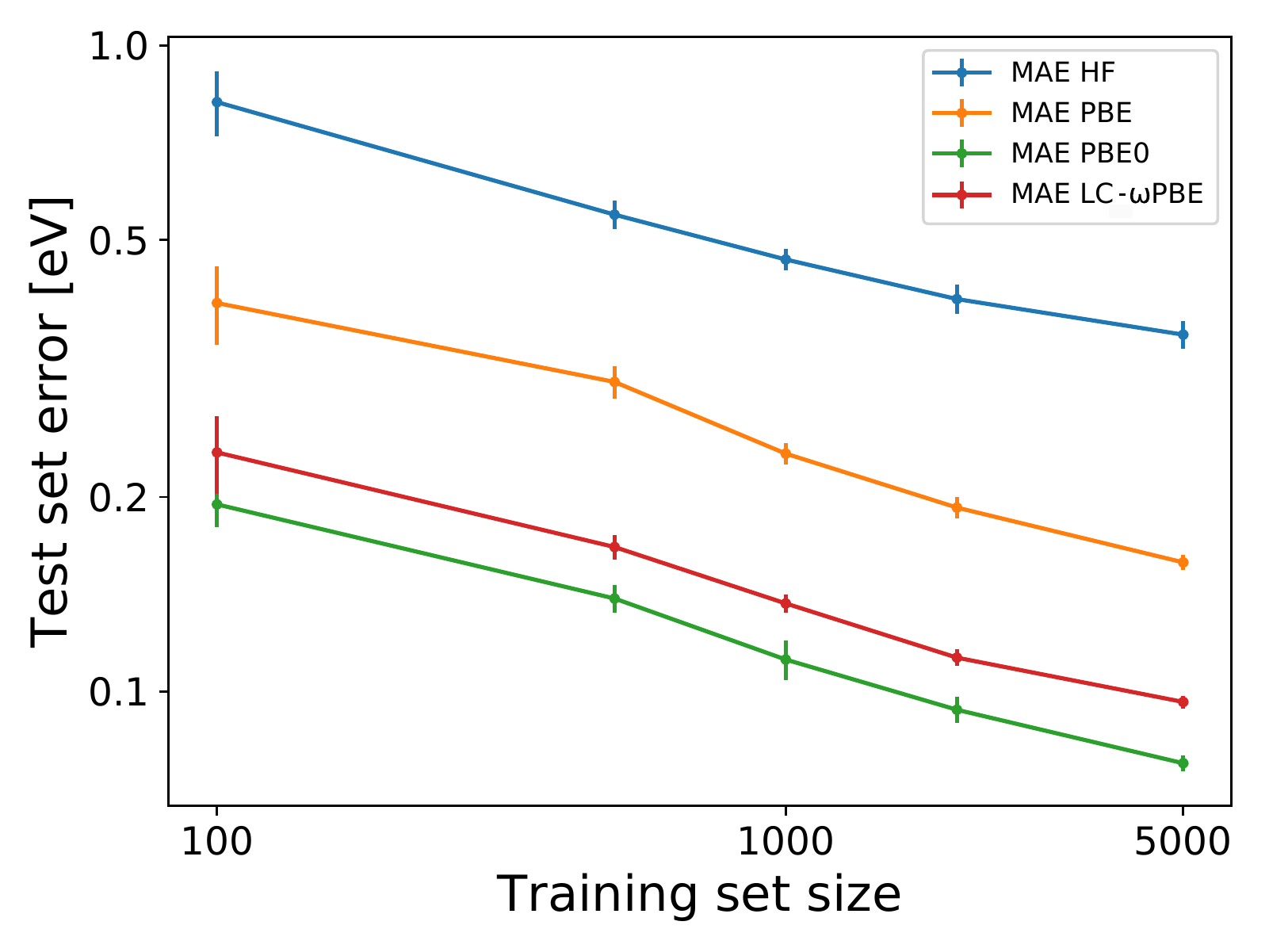}
    \caption{Learning curves of the average energy curvature ($C^N_{avg}$) in function of the training set size. The learning exercise is reported for three functionals and Hartree-Fock (HF) using the def2-SVP basis set. The error bars correspond to the standard deviation of the 10-fold cross-validation. The models was built using the SLATM molecular representation.}
    \label{fig:LearningCurve}
\end{figure}

The training set for the non-linear regression of $C^N_{avg}$ was selected by randomly choosing 6465 molecules out of the QM7 database, leaving the remaining 10\% for out-of-sample predictions. The performance of the model was then evaluated by training on 5 sub-sets of different sizes (100, 500, 1000, 2000 and 5000 molecules) while predicting on a validation set of fixed size (645 molecules). The final learning curve (Figure \ref{fig:LearningCurve}) is obtained by randomly sampling the training and the validation set 10 times and averaging the mean absolute errors (10-fold cross-validation). The regression model reported in the Figure uses the spectrum of London and Axilrod-Teller-Muto (SLATM) molecular representation,\cite{Huang2016,Huang2017} as it was the best performing for the largest training set size (further details about the global and local framework used herein are given in the Supplementary Material).

\begin{figure}[!htb]
    \centering
    \includegraphics[width=0.45\textwidth]{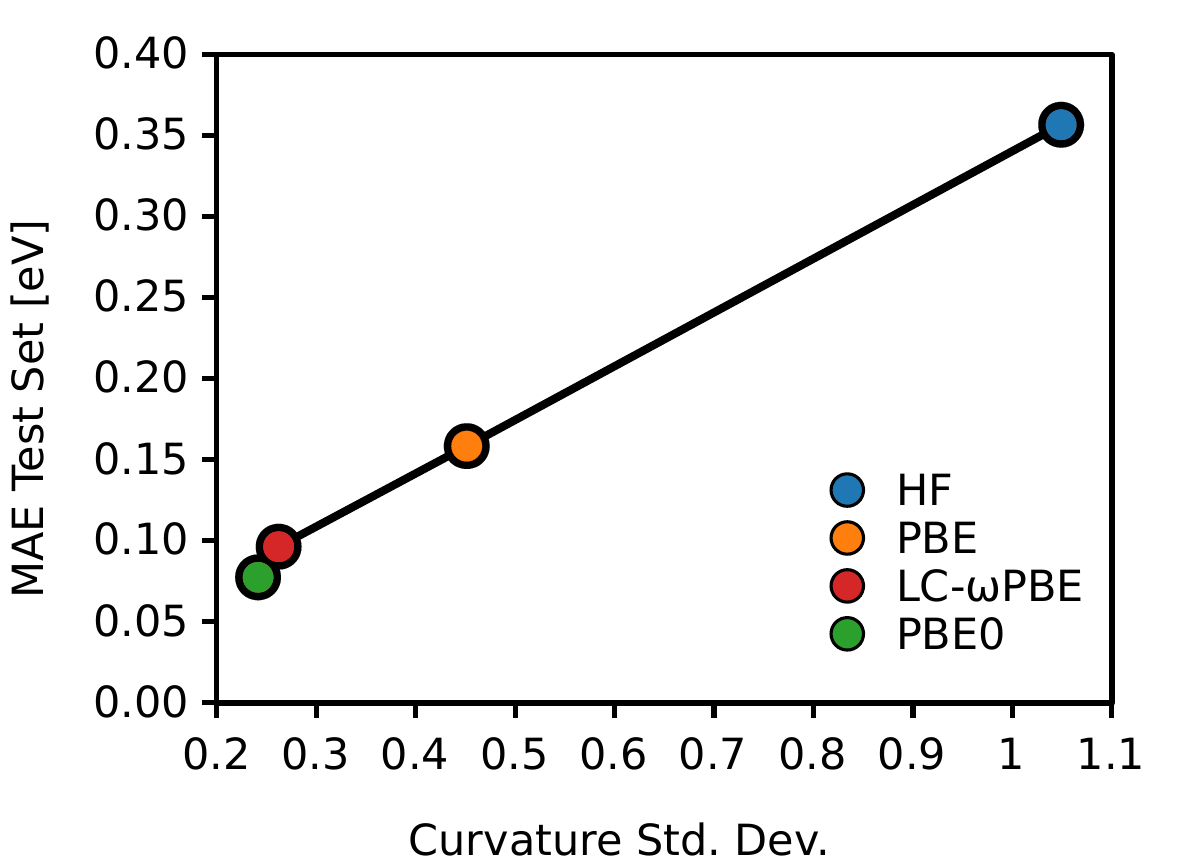}
    \caption{MAE of the model at a training set size of 5000 molecules in function of the standard deviation of $C^N_{avg}$ using the three functionals and HF.}
    \label{fig:Curvature}
\end{figure}

As shown in Figure \ref{fig:LearningCurve}, the difficulty of the learning exercise largely depends on the electronic structure level at which the energy curvature is computed. The learning of PBE0\cite{Adamo1999,Ernzerhof1999}/def2-SVP and LC-$\omega$PBE\cite{Vydrov2006,Vydrov2006a,Vydrov2007}/def2-SVP is the most straightforward, followed by PBE\cite{Perdew1996,Perdew1997}/def2-SVP and finally Hartree-Fock (HF/def2-SVP). This specific ordering is directly related to the amount of variation of the target quantity ($C^N_{avg}$) within each functional and method. As shown in Figure \ref{fig:Curvature}, the mean absolute error of the model trained on 5000 molecules correlates nearly perfectly with the standard deviation of $C^N_{avg}$ for each level of theory.

\begin{figure}[!htb]
    \centering
    \includegraphics[width=0.35\textwidth]{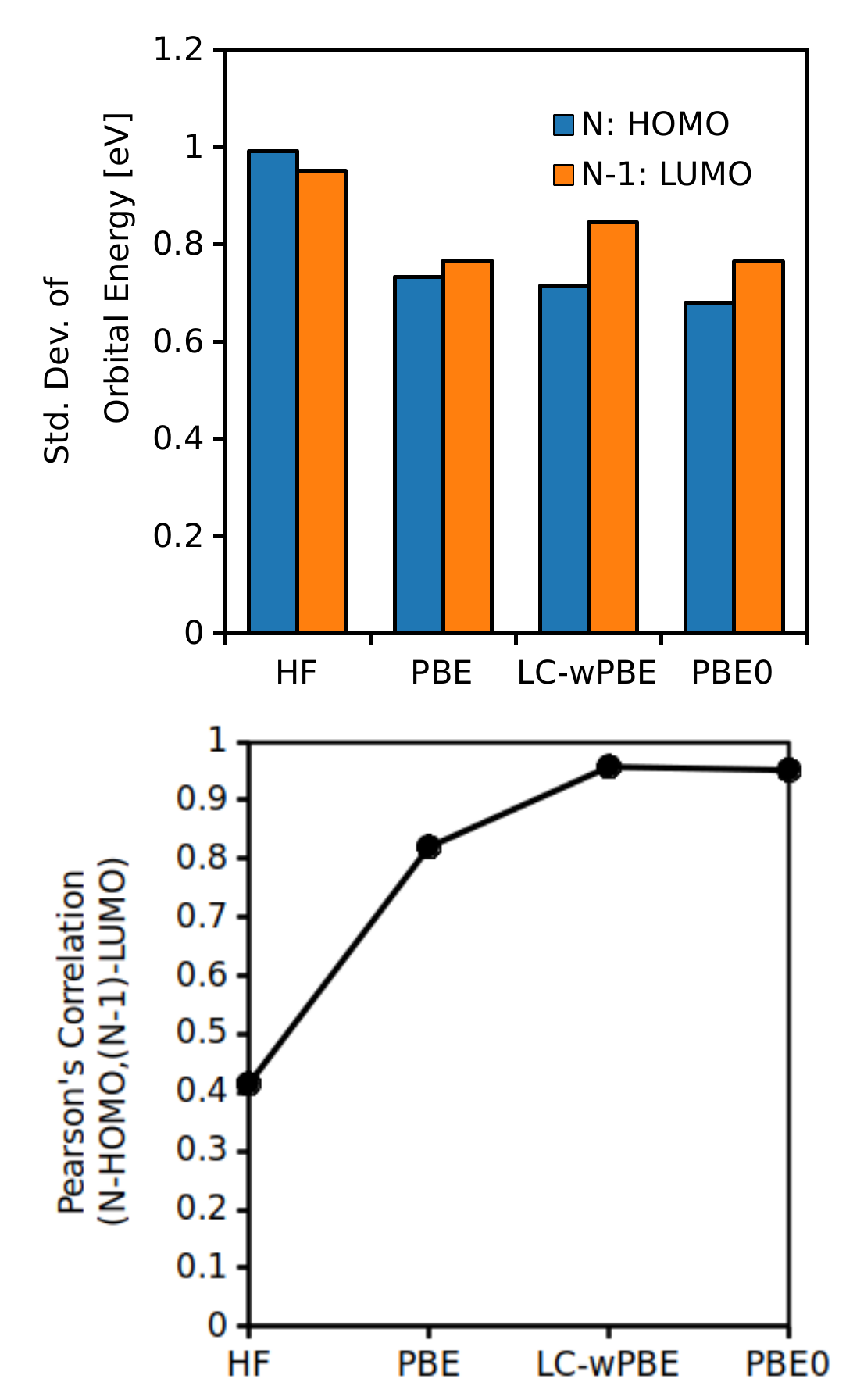}
    \caption{\emph{(top)} Standard deviations of N-HOMO and (N-1)-LUMO through QM7 with three functionals and HF. \emph{(bottom)} Pearson's correlation coefficient between the N-HOMO and (N-1)-LUMO energies at different levels of theory.}
    \label{fig:Variance}
\end{figure}

Following Eq. \ref{eq:curve}, the energy curvature depends on the HOMO eigenvalue of the neutral molecule [N-HOMO] and on the LUMO eigenvalue of its radical cation [(N-1)-LUMO]. Therefore, the relative robustness of each functional in describing these two quantities could be invoked to rationalize the overall spread of its $C^N_{avg}$. However, as shown in the top panel of Figure \ref{fig:Variance}, the individual variations of the frontier orbital energies are not sufficient to explain the overall trend found for $C^N_{avg}$. All the functionals are characterized by similar orbital energies standard deviations, whereas HF shows larger deviations. 

The ordering of Figure \ref{fig:Curvature} is retrieved only after combining the information about the variation of the orbital energies with the one about their correlation (Figure \ref{fig:Variance}, bottom panel). In particular, the frontier orbital eigenvalues correlate almost perfectly in LC-$\omega$PBE and PBE0, while their correlation is lower in PBE and very poor within HF. Consequently, the difficulty of the learning exercise ultimately depends on the consistency of a method in describing the orbital energies both of the neutral and the radical cation state of a molecule. Both the spread in orbital eigenvalues and their covariance do not change while performing the computation with a larger basis set (see Supplementary Material)

The poor covariance between the frontier orbital eigenvalues in Hartree-Fock is the consequence of the different way in which the occupied and the unoccupied manifolds are treated within the method. In particular, the orbital energies of the occupied manifold, hence the HOMO eigenvalue of the neutral molecule, is determined in Hartree-Fock by the effective potential of N-1 particles, as the exchange cancels out the self-interaction contribution. This is not the case for the unoccupied manifold, where the effective potential originates from the totality of the particles. In contrast, the energies of both the occupied and the unoccupied orbitals in density functional theory are determined by a N-1 particle effective potential, as the (approximate) exchange-correlation hole exclude a single electron from each and every orbital.

\section{\label{sec:level3} System specific $\gamma$-tuning}

The energy curvature predicted for each molecule by the machine-learning model can be readily applied as a criterion for system-specific $\gamma$-tuning of range-separated hybrid density functionals. Usually, the tuning procedure consists in adjusting the range-separation parameter to satisfy as closely as possible the Koopmans' theorem for both the neutral and the anionic state of a targeted molecule.\cite{Baer2010} While commonly used, this method requires the ionization potential to be known in advance, which limits, at best, the applicability of the procedure. As already demonstrated by Kronick, Baer and coworkers\cite{Stein2012}, the minimization of $C^N_{avg}$ in approximate functionals implies their compliance to the Koopmans' theorem. Therefore, the optimal range-separation parameter for a specific compound can be found by imposing the curvature to be identically zero.

\begin{figure}[!htb]
    \centering
    \includegraphics[width=0.45\textwidth]{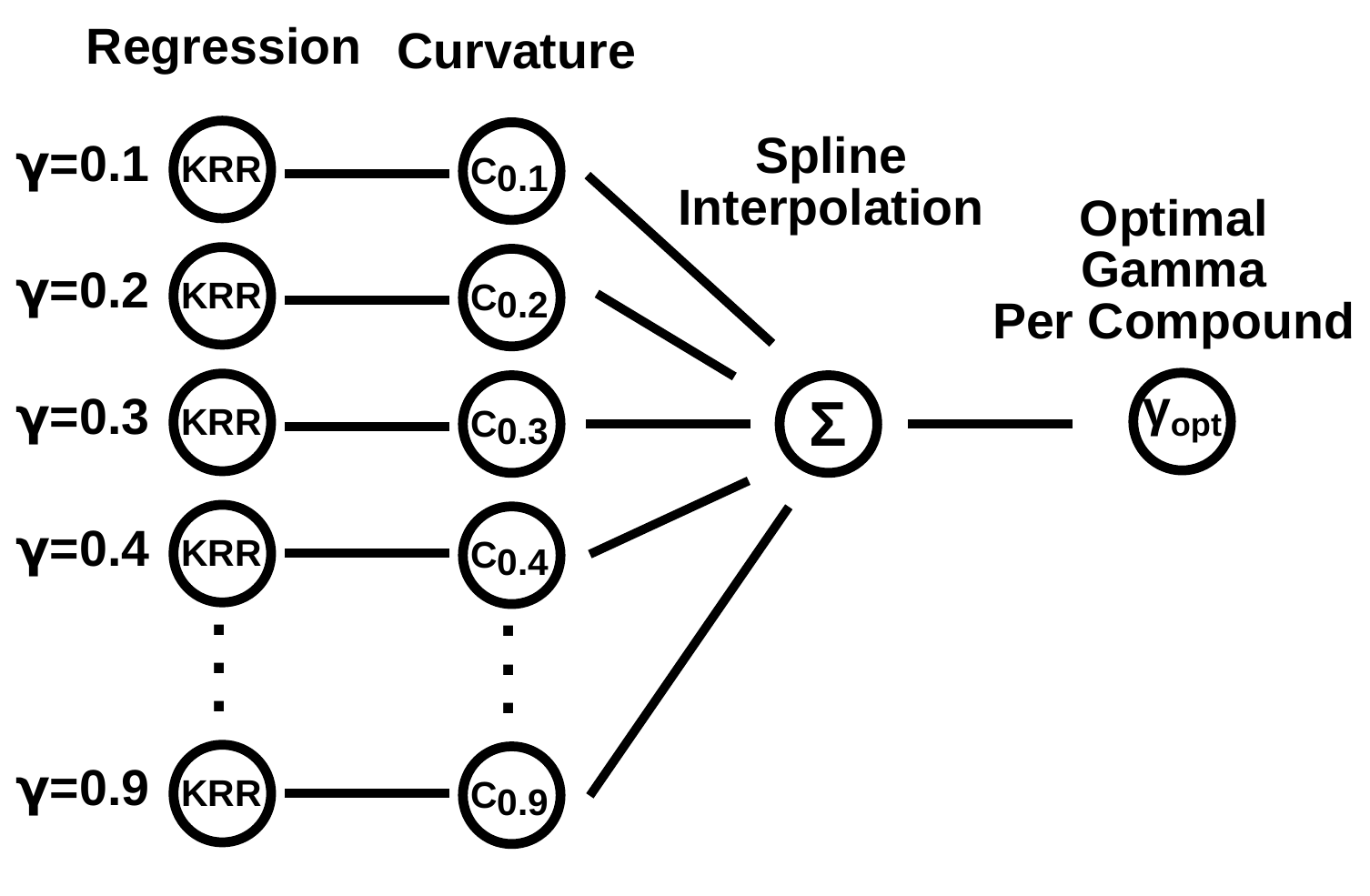}
    \caption{Schematic representation of the regression framework for the prediction of the optimal range separation parameter per compound. For each molecule, nine independent models predict the energy curvature at LC-$\omega$PBE at nine $\gamma$ values ranging from 0.1 to 0.9 Bohr$^{-1}$. The system-specific optimal $\gamma$ parameter, for which $C^N_{avg}=0$, is then found by a cubic spline interpolation.}
    \label{fig:Scheme}
\end{figure}

Figure \ref{fig:Scheme} schematically illustrates a modification of the regression framework as presented in the previous section, which uses the curvature information to determine the optimal $\gamma$ parameter for a given chemical system. In particular, the procedure consists in nine independent kernel ridge regression models, each targeting $C^N_{avg}$ at different values of the range-separation parameter. In the last step, a cubic spline interpolation of the predicted curvatures leads to the optimal $\gamma$ parameter (\textit{i.e.}, the $\gamma$ value for which $C^N_{avg}=0$) for a given molecule.

\begin{figure}[!t]
    \centering
    \includegraphics[width=0.45\textwidth]{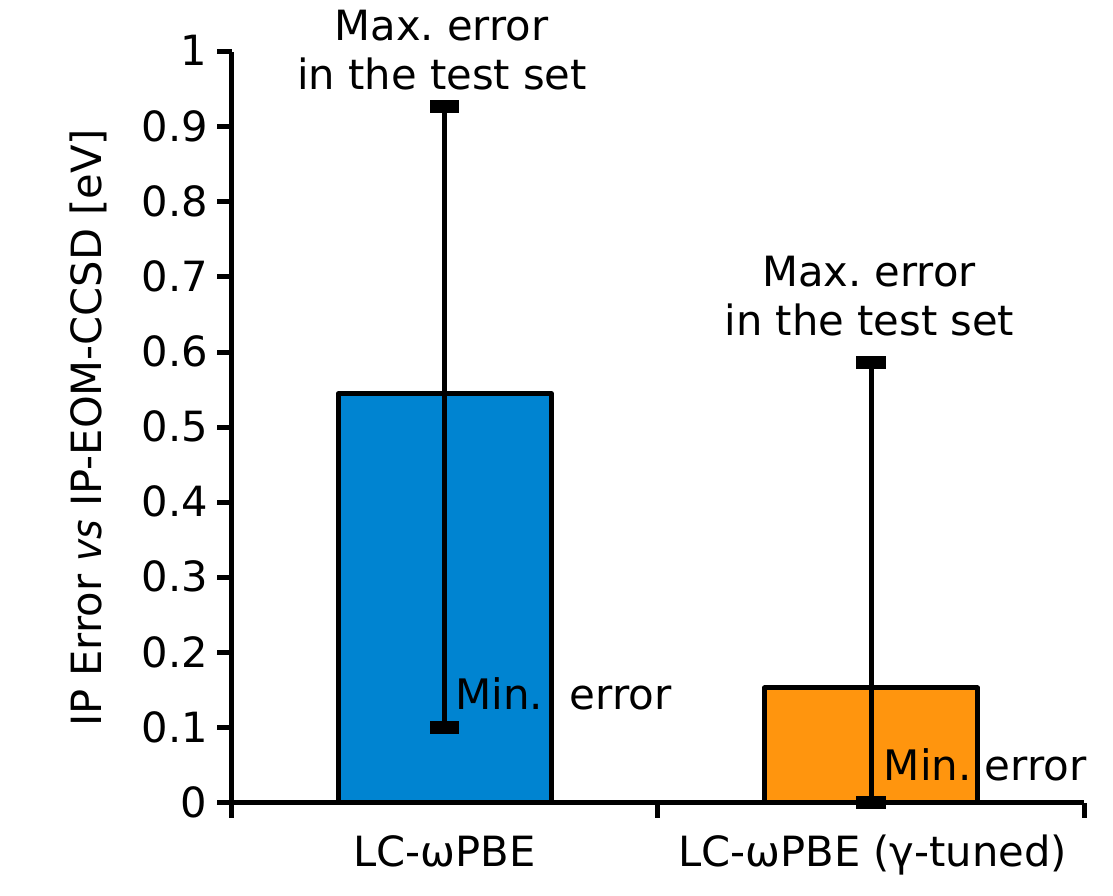}
    \caption{Absolute error between $-\epsilon_{HOMO}$ of LC-$\omega$PBE and its $\gamma$-tuned variant and the ionization potential at IP-EOM-CCSD across the 640 molecules of the test set. Optimal $\gamma$ values derive from the model described in Figure \ref{fig:Scheme}. The height of the histogram represents the mean absolute error, while the bars show the maximum and the minimum deviations.}
    \label{fig:IPError}
\end{figure}

To avoid the introduction of unpredictable noise in the data, we considered here only those compounds, for which all the computations converged. In consequence, the model was trained using the energy curvature of 5754 small organic molecules taken from the QM7 database and used to predict the system-specific optimal LC-$\omega$PBE $\gamma$ values for a test set of 640 molecules. Upon a single point computation using the tuned functional, the ionization potential of each molecule is evaluated as $-\epsilon_{HOMO}$ and compared to the corresponding value at IP-EOM-CCSD. For consistency, all computations are performed with the def2-SVP basis set. Figure \ref{fig:IPError} shows the accuracy of estimated IPs averaged over the test set for the standard LC-$\omega$PBE and its $\gamma$-tuned variant. The error bars show the maximum and the minimum deviation from the IP-EOM-CCSD reference registered among the 640 molecules.

The tuning procedure on the basis of the predicted curvatures results in a five-fold decrease of the average ionization potential error compared to the standard functional. The robustness of the predictions is further demonstrated by the fact that the worst error made with the $\gamma$-tuned variant is only as high as the average error made with the standard LC-$\omega$PBE.

\begin{figure}[!hbt]
    \centering
    \includegraphics[width=0.45\textwidth]{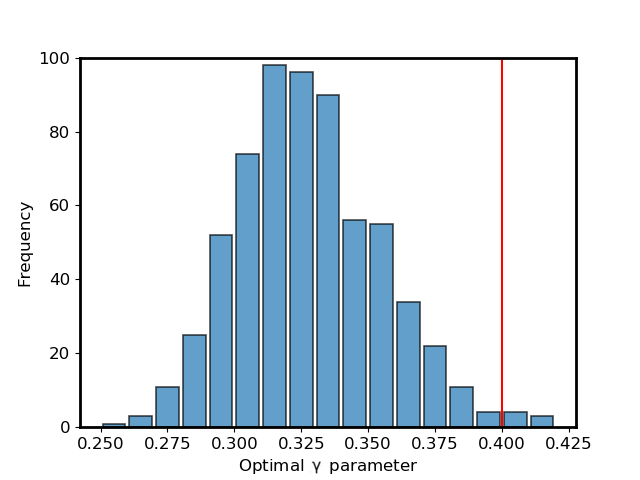}
    \caption{Distribution of the optimal $\gamma$ parameters [Bohr$^{-1}$] across the 640 molecules of the test set as predicted by the model described in Figure \ref{fig:Scheme}. The red line show the value of the range-separation parameter in the standard LC-$\omega$PBE.}
    \label{fig:Distribution}
\end{figure}

Including several hundreds of different molecules, the test set represents a sufficiently large ensemble for a statistically relevant analysis of the optimal range-separation parameter in LC-$\omega$PBE. By registering the frequency of appearance of the predicted $\gamma$ values, it is shown that their distribution tends to a Gaussian function centered around 0.32 Bohr$^{-1}$ (Figure \ref{fig:Distribution}). Out of the 640 molecules only 12 are characterized by an optimal $\gamma$ parameter close to the 0.4 Bohr$^{-1}$ of the standard functional. In all those cases where system specific $\gamma$-tuning is not possible, for instance in the computation of dimer binding energies,\cite{Karolewski2013} the distribution in Figure \ref{fig:Distribution} demonstrates that fixing the range-separation parameter of LC-$\omega$PBE to 0.32 Bohr$^{-1}$ would reduce the curvature for the majority of molecules.

The discrepancy with the original parametrization of LC-$\omega$PBE has to be interpreted as the results of a different optimization strategy. Here, the suggested 0.32 Bohr$^{-1}$ minimizes the energy curvature for highest number of compounds in a comprehensive dataset of organic molecules. Following the works of Baer,\cite{Stein2012,Baer2010} fixing the $\gamma$ parameter by minimization of the energy curvature is a formally motivated procedure, as it leads to compliance with the Koopmans' theorem and exact conditions of DFT. The original approach used for the parametrization of LC-$\omega$PBE is more pragmatic and seeks to minimize the error of the functional against different energy-based benchmark databases.\cite{Vydrov2006} The formal issue associated with this second strategy is that the range-separation parameter inevitably compensates for unrelated deficiencies in the rest of the approximated exchange-correlation functional.

\section{\label{sec:level3bis} Extrapolation}

The machine-learning models presented in the previous paragraphs rely on a global molecular representation, \textit{i.e.}, each vector in the feature space characterizes one specific compound. As the energy curvature is a molecular property, this class of representations is highly suitable and easily applicable to the regression problem. On the other hand, a model based on a global representation is not transferable, \textit{i.e.}, it cannot be trained on smaller compounds and used to make predictions on larger molecules.\cite{Huang_2018} This issue can be tackled using local representations, which encode the molecular information as a collection of atoms in their environments. By establishing similarity measures between local atomic environments, rather than between whole molecules, local representations lead to more transferable models, applicable to larger and more diverse molecules than those included in the training set (see, for instance, Refs. [\citenum{Fias2017,Unke2018,Fabrizio-CHEMSCI-2019}]). The regression framework shown in Figure \ref{fig:Scheme} is general and can be readily extended to local, atom-centered molecular representations. More details about the modification of the learning framework to accommodate locality and transferability are given in the Supplementary Material.

\begin{figure}[!htb]
    \centering
    \includegraphics[width=0.45\textwidth]{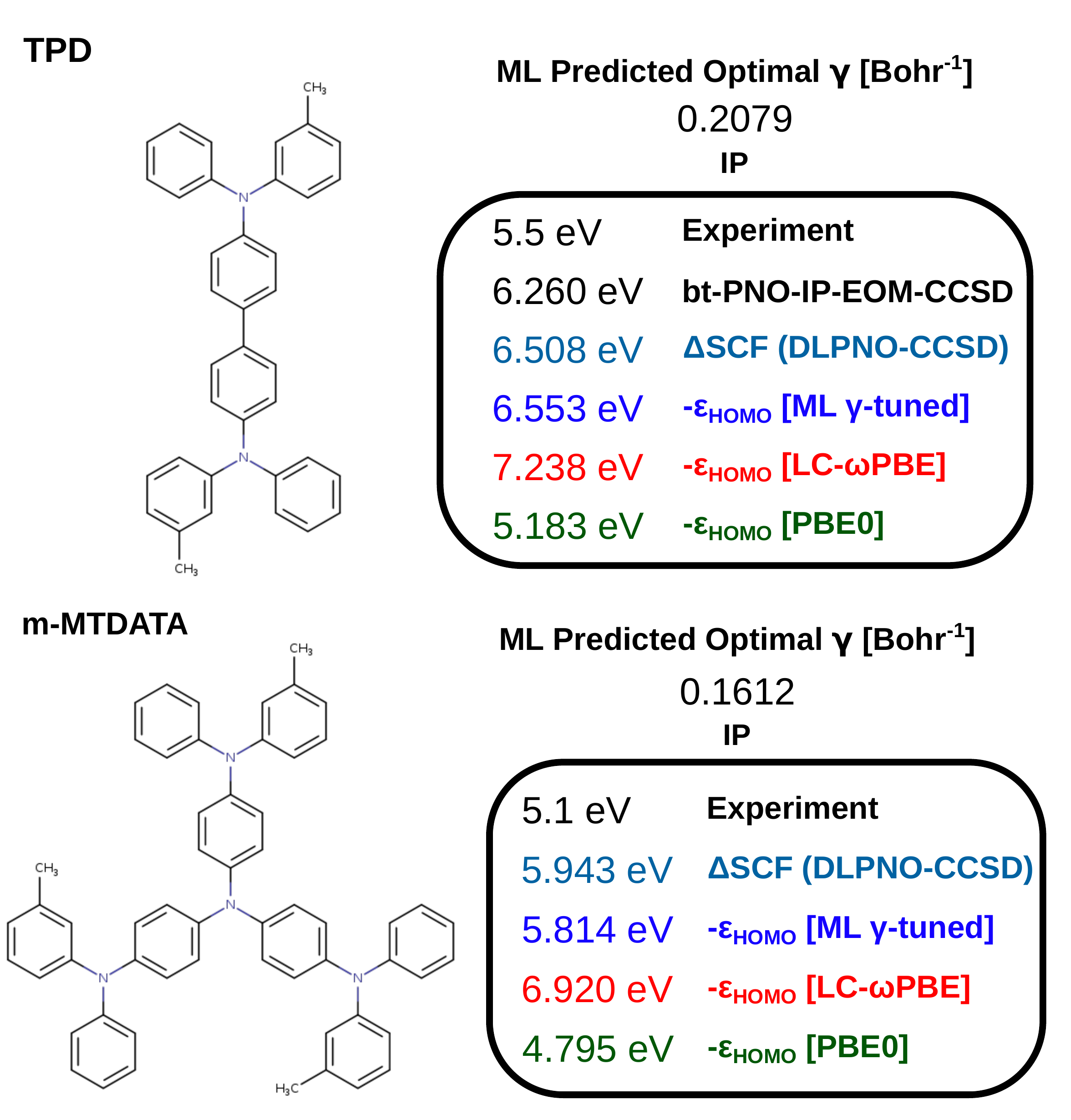}
    \caption{ Extrapolation: optimal $\gamma$ parameter derived from the model described in Figure \ref{fig:Scheme} for two large molecules relevant for the field of hole-transporting materials. The value of $-\epsilon_{HOMO}$ for both the standard LC-$\omega$PBE and its $\gamma$-tuned variant are reported along with reference ionization potentials. Experimental IPs are taken from Refs. [\citenum{Okumoto2001,Deotare2015}].}
    \label{fig:Extrapolation}
\end{figure}

Figure \ref{fig:Extrapolation} shows the application of the local regression framework to predict the optimal $\gamma$ values of two large molecules commonly used in hole-transporting materials:\cite{Jou2015,Shahnawaz2019} N,N'-Bis(3-methylphenyl)-N,N'-diphenylbenzidine (\textbf{TPD}) and 4,4',4''-Tris[(3-methylphenyl)phenylamino]triphenylamine (\textbf{m-MTDATA}). The model was exclusively trained on the local environments of the small organic molecules of the QM7 database using the atomic spectrum of London and Axilrod-Teller-Muto (aSLATM)\cite{Huang2016,Huang2017} representation combined with an orthogonal matching pursuit\cite{Mallat1992,Rubinstein2008} algorithm for sparse regression (see Supplementary Material).

The $-\epsilon_{HOMO}$ computed with standard LC-$\omega$PBE is a rather poor approximation of the ionization potential of \textbf{TPD} and \textbf{m-MTDATA} with errors around 1 eV compared to the \textit{ab-initio} references (bt-PNO-IP-EOM-CCSD and $\Delta$SCF at DLPNO-CCSD). Upon ML-based $\gamma$-tuning the error with respect the wavefunction based methods is reduced to 0.1-0.2 eV for both molecules. This result, obtained on compounds four time larger than the largest molecule in the training set, demonstrate the transferability of the local model and its applicability to targeted complex molecules. Interestingly, the optimal $\gamma$ parameters for both \textbf{TPD} and \textbf{m-MTDATA} are much lower than any value obtained on the smaller molecules of the QM7 test set (Figure \ref{fig:Distribution}). This behavior is consistent with the results of the existing literature\cite{Korzdorfer2011,Refaely-Abramson2011,Sun2013,Jacquemin2014} and further support the conclusion that $\gamma$ can be interpreted as the inverse of an effective conjugation length dependent on the system size. Finally, the HOMO eigenvalue of PBE0 is the farthest from the \textit{ab-initio} reference, but the closest to the experimental values. This result is not unexpected (see, for instance Refs. \citenum{Chi2016,Chi2016a}) and shows that the error made by the global hybrid mimics the effects of the condensed phase environment (\textit{e.g.} solvent, crystal field).\cite{Ghosh2011,Kotadiya2018}

\section{\label{sec:level4}Unsupervised learning and analysis of the QM7 dataset}

\begin{figure*}[!htb]
    \centering
    \includegraphics[width=0.875\textwidth]{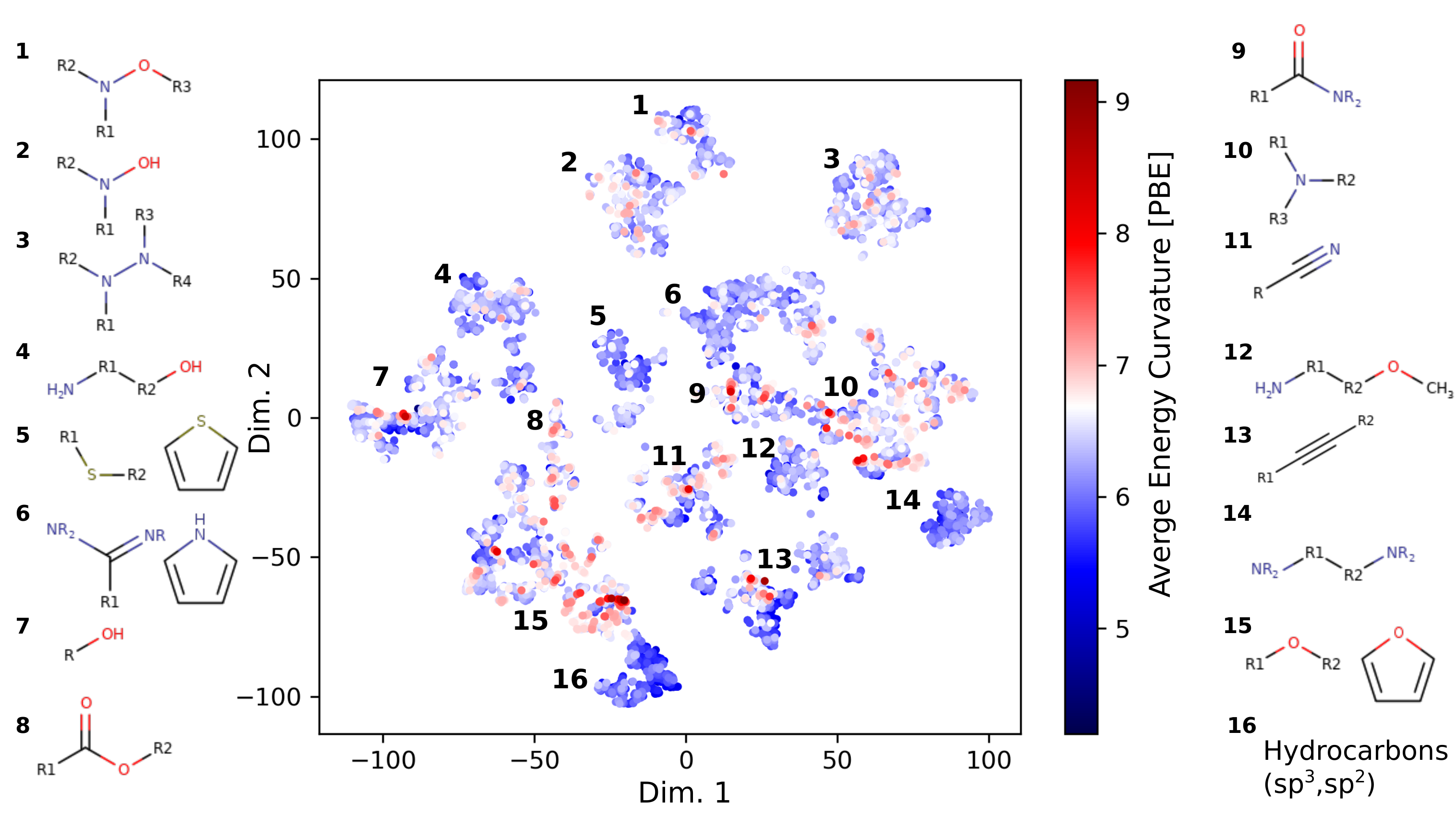}
    \caption{Two-dimensional t-SNE map of the QM7 database on the basis of the SLATM representation. Each point represents a compound colored by its average energy curvature computed with PBE/def2-SVP. The diverging color map highlights the data with the highest and the lowest average energy curvature. Each of the clusters contains molecules with similar patterns, which are defined by the corresponding numbering.}
    \label{fig:TSNE}
\end{figure*}

The large chemical diversity contained in the QM7 database promotes a thorough assessment of the relation between the energy curvature computed with a given functional and the system-specific structural and compositional patterns. However, drawing such a relationship for thousands of molecules inevitably leads to a high-dimensional problem, which is unsuitable for analysis and visualization. In this context, non-linear dimensionality reduction algorithms reveal the underlying structure of high-dimensional data by projecting complex vectors into lower dimensions.\cite{Das2006,Nguyen2006,Brown2008,Ferguson2010,Spiwok2011,Ferguson2011,Ceriotti2011,Tribello2012,Ceriotti2013,De2016,Reutlinger2012,Virshup2012,Duan2013,Schneider2017,Lemke2019} Figure \ref{fig:TSNE} shows a two-dimensional representation of the chemical diversity of the database using t-distributed Stochastic Neighbor Embedding (t-SNE).\cite{VanDerMaaten2008} This algorithm converts the similarity between molecules, which is defined herein as the euclidean distance between of their SLATM representation, to the probability of being each other's neighbors. The embedding of high-dimensional data into lower dimensions is then performed by ensuring that the joint probability between molecules should not change upon projection. While the two axis (dimensions) obtained after a t-SNE transformation have no formal physical or chemical meaning, it is still possible to identify at least a qualitative correlation between chemical properties and the dimensions in Figure \ref{fig:TSNE} \textit{vide infra}. 

The application of t-SNE to QM7 reveals clusters of compounds with similar chemical patterns, mainly defined by the presence or the absence of heteroatoms and their connectivity. In particular, the vertical axis (Dim. 2) somehow correlates with the number of heteroatoms, from zero (alkanes, bottom) to two or more non-carbon atoms (hydroxyamines and oxyamines, top). The horizontal axis follows instead a gradient of chemical composition going from the oxygen-based compounds (left) to nitrogen containing molecules (right), passing from mixed species. Each point is color-coded by its average energy curvature computed at PBE/def2-SVP to establish a global, qualitative connection between these macro-families of compounds and the degree of their deviation from piecewise linearity. The choice of PBE is motivated by the fact that the absence of Hartree-Fock exchange leads to a curvature that represents an upper limit for the other functionals.

Figure \ref{fig:TSNE} highlights seven key families characterized by at least one region of high average energy curvature (in red). Out of those clusters, three contains only oxygen as heteroatom (alcohols[7], ethers[15] and acids/esters [8]), two includes \textit{sp}-hybridized carbons (cyano groups [11] and alkynes [13]), one contains only nitrogen (amines [10]) and the last group includes the amides [9]. In contrast, alkanes (with the exception of the smallest methane and ethane, see discussion below)[16], diamines separated by long carbon chains [14], all sulfur containing compounds [5] and amidines [6] are all characterized by lower curvatures. These trends suggests that the presence of increasingly electron-rich heteroatoms tend to increase the average energy curvature. In particular, the presence of oxygen atoms is especially sensitive as shown by the qualitative difference between amides and amidines. The low average energy curvature that characterizes all sulfur containing compounds suggests that the presence of heteroatoms beyond the second row of the periodic table does not have a critical impact on the deviation from piecewise linearity. In addition to heteroatoms, the hybridization of the carbon centers is also a relevant factor as illustrated by the contrast between alkanes and alkynes groups. These conclusions are consistent with previous work on charge transfer complexes\cite{Ruiz1995,Ruiz1996} and delocalization error.\cite{Gani2016} In particular, the results presented here are comparable with work of Kronik and Baer,\cite{Stein2012} who report the average energy curvature for a set of nine small molecules, whose order can be rationalized in terms of the presence of electron-rich heteroatoms, their hybridization and the molecular size. 

\begin{figure}[!htb]
    \centering
    \includegraphics[width=0.425\textwidth]{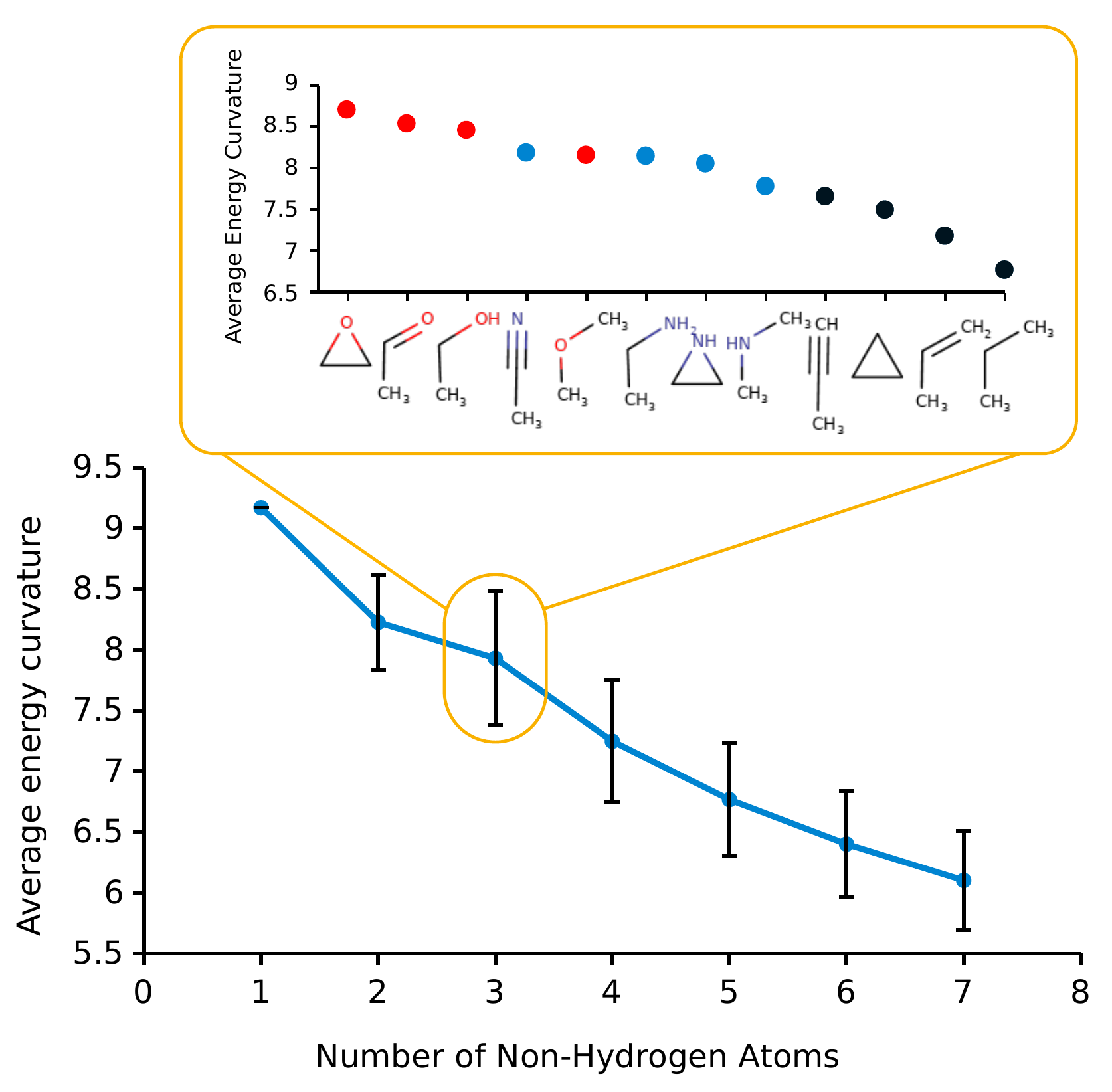} 
    \caption{Correlation between the average energy curvature at PBE/def2-SVP and the size of the molecules. The mean values (dots) are computed averaging $C^N_{avg}$ over all the compounds with the same number of non-hydrogen atoms. The error bar represents one standard deviation from the mean. The inset shows the average energy curvature of all the compounds in QM7 with 3 non-hydrogen atoms. The color code in the inset highlights the presence of oxygen (red), nitrogen (blue) or carbon only compounds (black).}
    \label{fig:correlation}
\end{figure}

Although not explicitly evident in the mapping of Figure \ref{fig:TSNE}, the molecular size is in fact crucial to determine the extent of the average energy curvature. To emphasize this point, Figure \ref{fig:correlation} correlate the curvature at PBE/def2-SVP and the size of the molecules, upon averaging $C^N_{avg}$ over all the molecules with the same number of non-hydrogen atoms ($N_{heavy}$). Although the mean values for $N_{heavy}=1$ and $N_{heavy}=2$ are not statistically significant (\textit{i.e.}, these categories include only 1 and 3 molecules respectively), the robust inverse size/curvature relationship justifies the high curvature of the smallest alkanes (Cluster 16 Figure \ref{fig:TSNE}). 

The error bars in Figure \ref{fig:correlation} shows that within every $N_{heavy}$ there is a distribution of curvatures that reflect the chemical composition. The analysis of the subset with 3 non-hydrogen atoms is especially suitable as it contains sufficient compounds to reflect general trends but is simultaneously small enough to list all its molecules. The inset of Figure \ref{fig:correlation} show the energy curvature of all the compounds with $N_{heavy}=3$ ordered from the highest to the lowest. This plot validate the conclusions drawn from the t-SNE map, as the curvature decreases with the electron-richness of the heteroatom (O > N > C). One exception due to the effects of hybridization is acetonitrile, which has a slightly higher, but comparable curvature to methyl ether. Complementing the information of the t-SNE map, the inset Figure reveals the high-energy curvature of 3-membered rings (oxirane, aziridine and cyclopropane) that are generally considered to act as unsaturated systems.\cite{Dewar1979} The previous arguments remain valid to explain the relative order of the electron rich 5-membered conjugated heterocycles [thiophene (6.3) < pyrrole (6.5) < furane (6.7)], whose curvatures are as expected higher than benzene (6.0).

\section{\label{sec:level5}Conclusion}

The average energy curvature with respect to the particle number is a crucial system-dependent property of density functionals, which quantify their deviation from the exact conditions of DFT and therefore affects their accuracy. Related to the lack of derivative discontinuity in the exchange-correlation potential and thus to the degree of severity of the delocalization error, the information about this quantity has been successfully used for optimal tuning of long-range corrected functionals and to correct Kohn-Sham orbital eigenvalues to match ionization potentials and electron affinities. In this work, we have proposed the construction of a machine-learning model of the average energy curvature and shown its applications for the system specific tuning of the LC-$\omega$PBE functional. In parallel, unsupervised learning techniques have been applied to obtain qualitative information about particular chemical patterns and molecular properties which results into highly convex curvatures. 

As the curvature is both a system specific and a functional dependent quantity, we have first shown that the learning exercise is not equally difficult for any given functional, but it depends on its ability to describe on equal footing the neutral and the radical cation state of a molecule. This result implies that the possible spread of value for the average energy curvature is not equal for all methods. In particular, the largest standard deviation for the curvature is registered for Hartree-Fock, due to the poor correlation between the neutral molecule HOMO eigenvalue and the LUMO of the radical cation. 

Training several independent models to target the curvature at LC-$\omega$PBE for different values of its range-separation parameter led to the construction of a second framework dedicated to the system-dependent optimal tuning of the functional. The use of the predicted $\gamma$ parameters resulted in a five-fold increase of the accuracy when estimating the first ionization potential (IP) with $-\epsilon_{HOMO}$ with respect to the standard functional. The distribution of the predicted range-separation parameters on the QM7 database shows that the original 0.4 value of LC-$\omega$PBE is far from optimal to minimize energy curvature. As a generalization of the framework, we use a local molecular representation for the training and demonstrate the transferability of the modified model by estimating the optimal $\gamma$-values and computing the ionization potentials of two larger molecules, relevant for the field of hole-transporting materials.

Finally, projecting the high dimensional SLATM representation of QM7 in two dimensions with a t-SNE algorithm revealed the underlying structure of the database. In particular, the mapping showed several distinct clusters enclosing molecules similar to each other in terms of their scaffold and presence of heteroatoms. The curvature values across these clusters resulted the highest for compounds with second row heteroatoms, most frequently oxygen, or for compounds with sp-hybridization. Additional analysis of the data supports the existence of an inverse correlation between molecular size and the average energy curvature. 

\begin{acknowledgments}
The authors acknowledge Kun-Han Lin for helpful discussions and the National Centre of Competence in Research (NCCR) "Materials' Revolution: Computational Design and Discovery of Novel Materials (MARVEL)" of the Swiss National Science Foundation (SNSF), as well as the EPFL for financial support.
\end{acknowledgments}

\section*{Data Availability Statement}

The data and the model that support the findings of this study are openly available in Materials Cloud.

\appendix

\section{Computational Details}

The molecular geometries for all species were taken as published in the QM7 database.\cite{Blum2009,Rupp2012} The curvatures were computed according to Equation \ref{eq:curve} using the orbital eigenvalues of the neutral and the first radical cation state of each molecule. All the computations using PBE, PBE0, LC-$\omega$PBE and Hartree-Fock were performed in Gaussian16,\cite{g16} in combination with the def2-SVP\cite{Weigend2005} basis set. The first ionization potential energies at IP-EOM-CCSD\cite{Sinha1989,Stanton1994} and bt-PNO-IP-EOM-CCSD\cite{Dutta2016,Dutta2018}, as well as at $\Delta$SCF (DLPNO-CCSD)\cite{Riplinger2013a} were obtained with Orca 4.0\cite{Neese2012} using the def2-SVP basis set for consistency with the DFT values. The density fitting approximation was applied in the DLPNO-CCSD and bt-PNO-IP-EOM-CCSD computations. The machine-learning representations and similarity kernels were obtained using the Quantum Machine Learning toolkit QMLcode\cite{Christensen2017} with the exception of SOAP\cite{Bartok2013} (see Supplementary Material), which was computed using DScribe 0.3.2.\cite{Himanen2020} The mathematical form of the similarity kernels was chosen as standard procedure according to the specific representation (more details in the Supplementary Material). The two-dimensional map of the QM7 database was generated using the t-SNE\cite{VanDerMaaten2008} algorithm as implemented in the scikit-learn package.\cite{scikit-learn}

\section*{References}
%\nocite{*}
\bibliography{bibliography}% Produces the bibliography via BibTeX.

\end{document}

\LARGE {\textbf{Supplementary Material}}

\vspace*{0.75cm}

\large \textbf{Learning the energy curvature versus particle number in approximate density functionals}

\vspace*{0.25cm}

Alberto Fabrizio,\textit{$^{a,b}$} Benjamin Meyer,\textit{$^{a,b}$} and Clemence Corminboeuf$^\ast$\textit{$^{a,b}$} \\

%Affiliations
\noindent\small{\textit{$^{a}$~Laboratory for Computational Molecular Design, Institute of Chemical Sciences and Engineering, \'Ecole Polytechnique F\'ed\'erale de Lausanne, CH-1015 Lausanne, Switzerland.\\$^{b}$~National Centre for Computational Design and Discovery of Novel Materials (MARVEL), {\'E}cole Polytechnique F{\'e}d{\'e}rale de Lausanne, 1015 Lausanne, Switzerland \\ ~\\ E-mail: clemence.corminboeuf@epfl.ch}}

\section{Performance of different molecular representations in learning the average energy curvature}

\begin{figure}[!htb]
    \centering
    \includegraphics[width=0.6\textwidth]{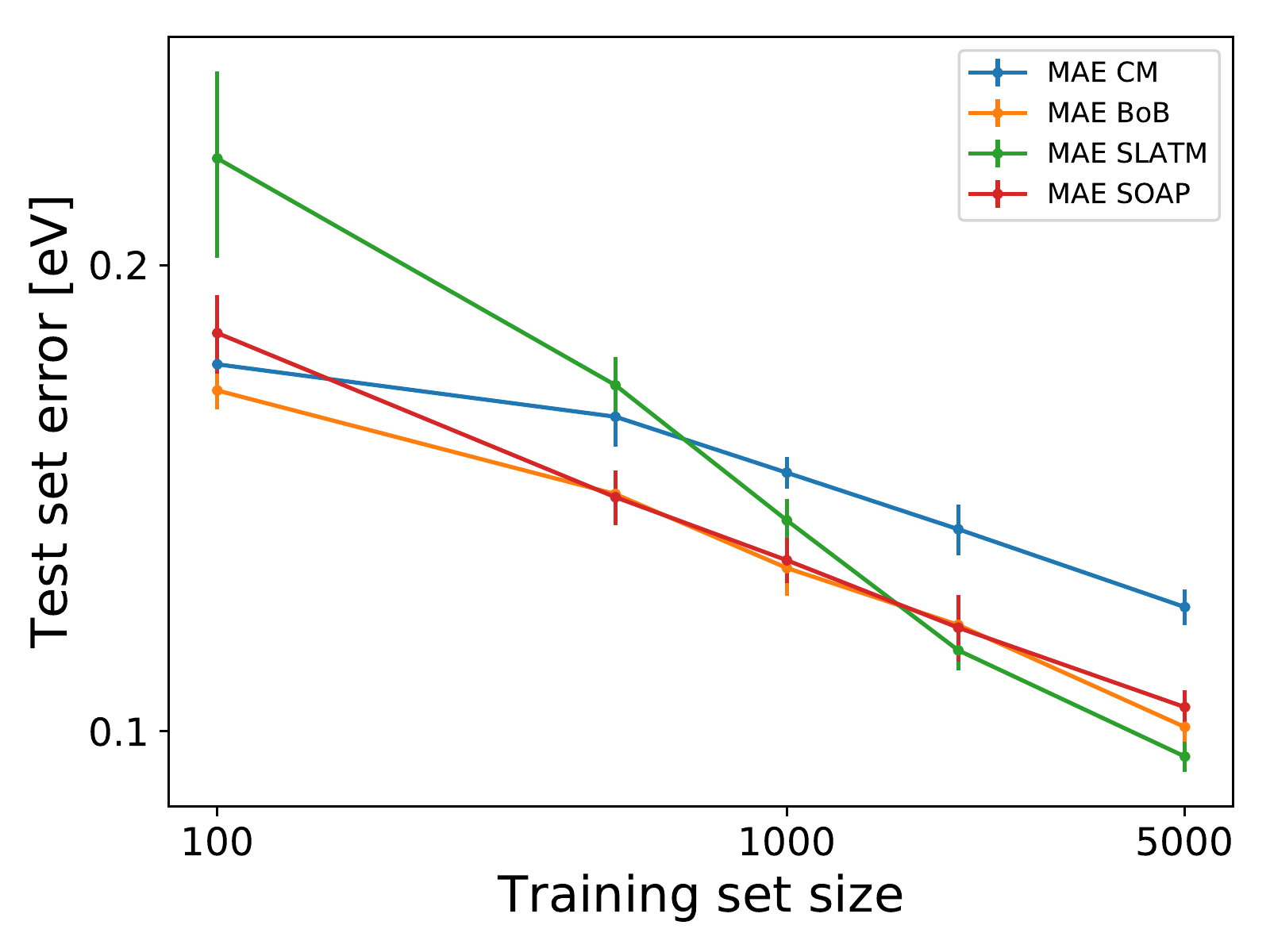}
    \caption{Learning curves of the average energy curvature ($C^N_{avg}$) at LC-$\omega$PBE/def2-SVP in function of the training set size. The learning exercise is reported for four different molecular representations: the Coulomb matrix (CM), the Bag of Bonds (BoB), the spectrum of London and Axilrod-Teller-Muto (SLATM) and the smooth overlap of atomic positions (SOAP). The error bars correspond to the standard deviation of the 10-fold cross-validation.}
    \label{fig:LearningCurveRepresentations}
\end{figure}

The performance of a machine-learning model targeting chemical properties depends strongly on the way the molecular information is represented.\cite{Huang2016,Bartok2013} A suitable representation constitutes in fact a meaningful relationship between the target property (herein the average energy curvature) and the molecular structure and composition. Over the last few years, several physically motivated molecular representations have been proposed, each of them including an increasing amount of chemical information.\cite{Behler2011,Rupp2012,Montavon2013,Bartok2013,Schutt2014,Hansen2015,Ghiringhelli2015,Faber2015,Botu2015,Faber2016,Huang2016,Huang2017} Figure \ref{fig:LearningCurveRepresentations} shows the performance in terms of mean absolute error of the average energy curvature computed at LC-$omega$PBE level using four different molecular representations: the Coulomb matrix (CM),\cite{Rupp2012} the Bag of Bonds (BoB),\cite{Hansen2015} the spectrum of London and Axilrod-Teller-Muto (SLATM)\cite{Huang2016,Huang2017} and the smooth overlap of atomic positions (SOAP).\cite{Bartok2013}

Overall, the SLATM representation leads to the lowest mean absolute error at the full training set and to the steepest learning curve. The final accuracy of the other representations tested is nevertheless comparable, resulting in particularly small deviations ranging from 4 meV (BoB) to 24 meV (CM). 

\section{Local framework for the regression of the energy curvature}

\begin{figure}[!htb]
    \centering
    \includegraphics[width=1\textwidth]{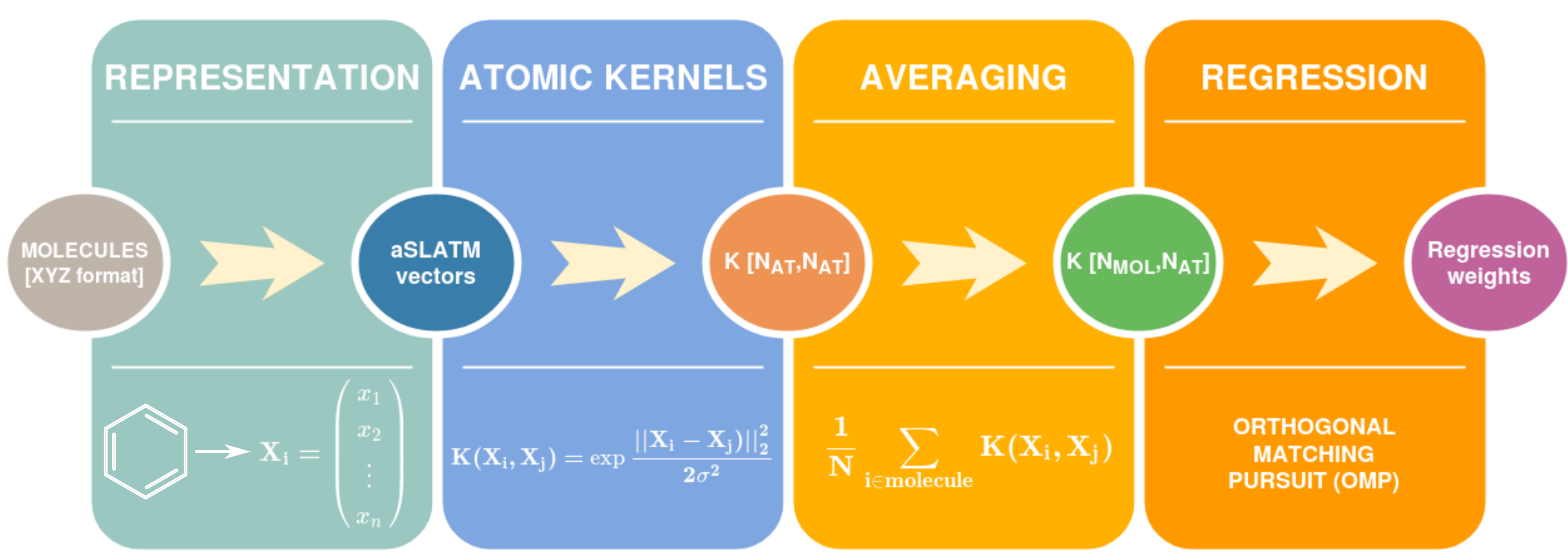}
    \caption{Schematic representation of the local framework for the regression of the energy curvature.}
    \label{fig:Scheme}
\end{figure}

The average energy curvature ($C_{avg}^N$) for a fixed functional is a molecular property, whose partitioning into atomic contributions cannot be defined uniquely. Instead of imposing \textit{a priori} a decomposition scheme, we construct a machine-learning model able to perform the regression and simultaneously find the most suitable atomic partitioning of $C_{avg}^N$. Figure \ref{fig:Scheme} is a schematic representation of such a regression framework. First, the molecular information is vectorized as a collection of atomic environments using the aSLATM representation. Then, Gaussian similarity kernels are evaluated between all the local environments, resulting in a $N_{at}~X~N_{at}$ matrix, where $N_{at}$ is the number of atoms in the training set. Since the dimensionality of the target $C_{avg}^N$ is instead equal to the number of compounds, the lines of the kernel matrix are averaged for the atoms belonging to each molecule. Building a molecular similarity measure by averaging its local atomic contributions is not a novelty, but it represents the most straightforwards solution when evaluating the similarity of different compounds on the basis of a local representation.\cite{De2016} 

The rectangular kernel resulting from the averaging procedure cannot be directly inverted to solve the regression problem. This over-complete (redundant) problem can be tackled using a sparse regression technique originally developed for signal recovery: the orthogonal matching pursuit (OMP) algorithm.\cite{Mallat1992,Rubinstein2008} Given a fixed number of non-zero parameters($n_{NonZeroCoeff}$), this method is able to approximate the optimum regression weights vector ($\omega_{sol}$) by 

\begin{equation}
    \omega_{sol} = arg~min ~||Y- K \omega||^2_2 ~~~\text{Subject to}~~~ ||\omega||_0 \leq n_{NonZeroCoeff}
\end{equation}

where K is a over-complete kernel and Y is the regression target. For the model presented in this work the 300 non-zero coefficients were found to be optimal.

\section{Basis set dependence}

\begin{figure}[!htb]
    \centering
    \includegraphics[width=0.75\textwidth]{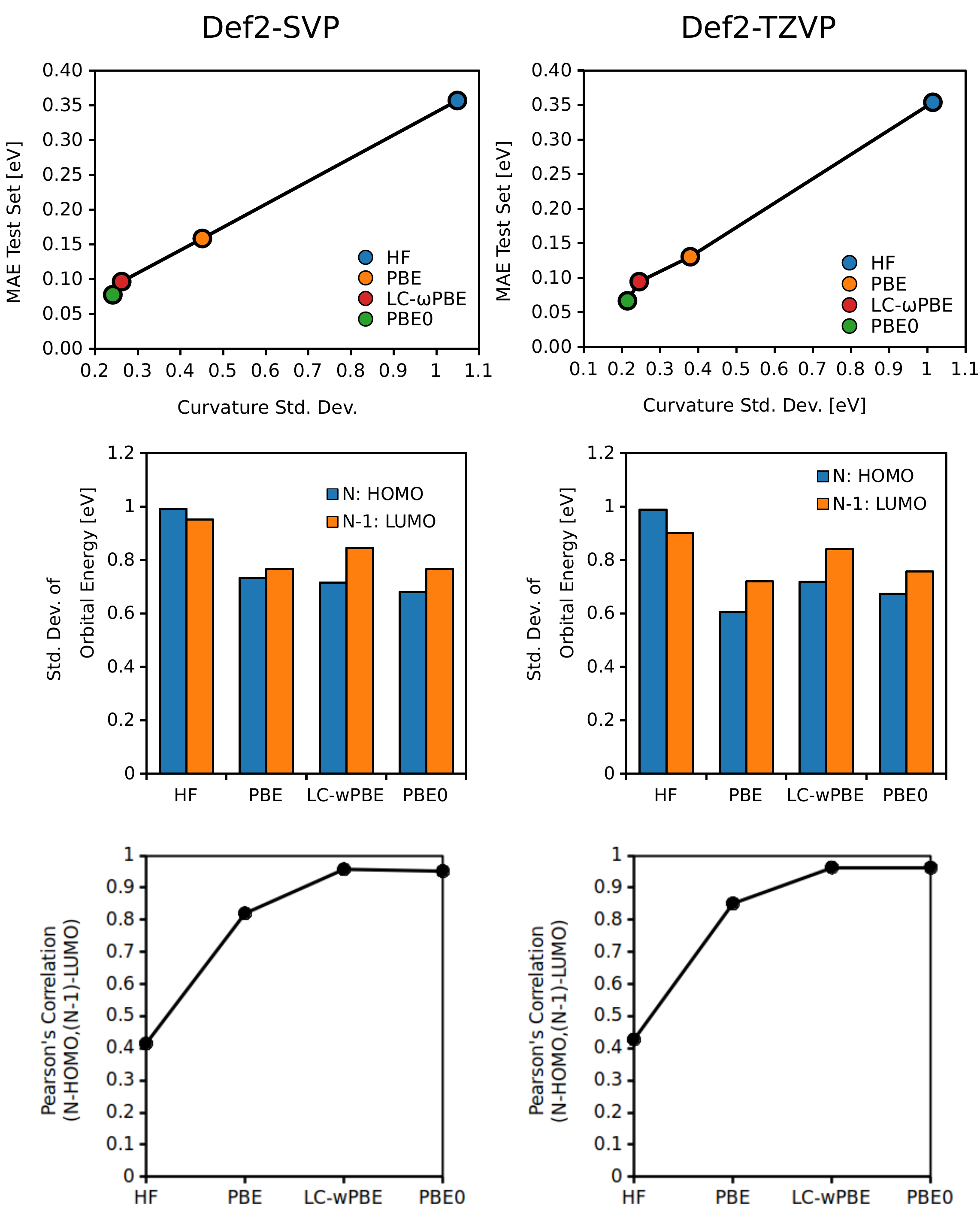}
    \caption{ (\textit{top}) MAE of the model at a training set size of 5000 molecules in function of the standard deviation of $C^N_{avg}$. (\textit{middle}) Standard deviations of N-HOMO and (N-1)-LUMO. (\textit{bottom}) Pearson's correlation coefficient between the N-HOMO and (N-1)-LUMO energies}
    \label{fig:BasisSet}
\end{figure}

The robustness of our conclusions with respect to the basis set used was tested by recomputing the MAE of the model for the full training set, as well as the standard deviation of the orbital eigenvalues and their correlation with def2-TZVP. As shown in Figure \ref{fig:BasisSet}, increasing the basis set size leads only to minor differences with respect to the results reported in the main text.

\newpage

\section{Numerical results}

The following tables contains the numerical data used in the Figures presented in this work.\newpage

\begin{longtable}{lm{2cm}m{1.5cm}m{3cm}m{1.7cm}m{3cm}}
\caption{Learning curves of $C^N_{avg}$: mean absolute error [MAE] and standard deviation of the 10-fold cross validation for each functional and Hartree-Fock using the SLATM representation. Additional details about the type of kernel and hyperparameters are added.}
\label{tab:LearningCurveMethods}\\\hline
Method & Kernel & Sigma & Training Size & MAE {[}eV{]} & Std. Dev. {[}eV{]} \\\hline
\endhead
HF & Gaussian & 207.5 & 100 & 0.817 & 0.094 \\
 &  &  & 500 & 0.547 & 0.028 \\
 &  &  & 1000 & 0.466 & 0.018 \\
 &  &  & 2000 & 0.405 & 0.021 \\
 &  &  & 5000 & 0.357 & 0.017 \\
PBE & Gaussian & 96.7 & 100 & 0.399 & 0.056 \\
 &  &  & 500 & 0.301 & 0.018 \\
 &  &  & 1000 & 0.233 & 0.009 \\
 &  &  & 2000 & 0.193 & 0.007 \\
 &  &  & 5000 & 0.158 & 0.004 \\
LC-$\omega$PBE & Gaussian & 107.8 & 100 & 0.234 & 0.032 \\
 &  &  & 500 & 0.167 & 0.007 \\
 &  &  & 1000 & 0.137 & 0.004 \\
 &  &  & 2000 & 0.113 & 0.003 \\
 &  &  & 5000 & 0.096 & 0.002 \\
PBE0 & Gaussian & 95.7 & 100 & 0.195 & 0.015 \\
 &  &  & 500 & 0.139 & 0.007 \\
 &  &  & 1000 & 0.112 & 0.008 \\
 &  &  & 2000 & 0.094 & 0.004 \\
 &  &  & 5000 & 0.077 & 0.002\\\hline
\end{longtable}

\begin{longtable}{lm{2cm}m{3cm}m{3cm}m{1.7cm}m{3cm}}
\caption{Learning curves of $C^N_{avg}$ at LC-$\omega$PBE: mean absolute error [MAE] and standard deviation of the 10-fold cross validation for each representation. Additional details about the type of kernel and hyperparameters are added.}
\label{tab:LearningCurveRepr}\\\hline
Representation & Kernel & Sigma & Training Size & MAE {[}eV{]} & Std. Dev. {[}eV{]} \\\hline
\endhead
CM & Laplacian & 196.9 & 100 & 0.173 & 0.006 \\
 &  &  & 500 & 0.16 & 0.007 \\
 &  &  & 1000 & 0.147 & 0.003 \\
 &  &  & 2000 & 0.135 & 0.005 \\
 &  &  & 5000 & 0.12 & 0.003 \\
BoB & Laplacian & 103.6 & 100 & 0.166 & 0.005 \\
 &  &  & 500 & 0.142 & 0.004 \\
 &  &  & 1000 & 0.127 & 0.005 \\
 &  &  & 2000 & 0.117 & 0.005 \\
 &  &  & 5000 & 0.101 & 0.002 \\
SLATM & Gaussian & 107.8 & 100 & 0.234 & 0.032 \\
 &  &  & 500 & 0.167 & 0.007 \\
 &  &  & 1000 & 0.137 & 0.004 \\
 &  &  & 2000 & 0.113 & 0.003 \\
 &  &  & 5000 & 0.096 & 0.002 \\
SOAP & Polynomial & 0.4 & 100 & 0.181 & 0.011 \\
 &  & $n_{max}$ = 8 & 500 & 0.142 & 0.006 \\
 &  & $l_{max}$ = 6 & 1000 & 0.129 & 0.004 \\
 &  & Cutoff = 4.0 \AA & 2000 & 0.117 & 0.006 \\
 &  &  & 5000 & 0.104 & 0.003\\\hline
\end{longtable}

\begin{longtable}{m{2cm}m{2cm}m{2cm}m{2cm}m{3cm}m{3cm}}
\caption{Standard deviation of the average energy curvature ($C^N_{avg}$), HOMO and LUMO orbital eigenvalues for each functional and HF. In addition the covariance of the frontier orbitals is reported along with the Pearson's correlation coefficient.}
\label{tab:stdsandCov}\\\hline
Functional & $\sigma(C^N_{avg})$ & $\sigma(\mathrm{HOMO})$ & $\sigma(\mathrm{LUMO})$ & HOMO-LUMO Covariance & Pearson's coefficient \\\hline
\endhead
HF & 1.049 & 0.992 & 0.952 & 0.392 & 0.415 \\
PBE & 0.451 & 0.733 & 0.767 & 0.461 & 0.82 \\
LC-$\omega$PBE & 0.262 & 0.716 & 0.846 & 0.58 & 0.957 \\
PBE0 & 0.241 & 0.68 & 0.766 & 0.496 & 0.951\\\hline
\end{longtable}

\begin{longtable}{m{4cm}m{3cm}m{3cm}m{3cm}}
\caption{Absolute errors between $-\epsilon_{HOMO}$ of LC-$\omega$PBE and its $\gamma$-tuned variant and the ionization potential at IP-EOM-CCSD across the 640 molecules of the test set.}
\label{tab:gamma}\\\hline
Functional & MAE {[}eV{]} & Min Err. {[}eV{]} & Max Err. {[}eV{]} \\\hline
\endhead
LC-$\omega$PBE & 0.545 & 0.100 & 0.927 \\
LC-$\omega$PBE ($\gamma$-tuned) & 0.153 & 0.000 & 0.586\\\hline
\end{longtable}
\newpage

\begin{longtable}{cccc}
\caption{Mean values and standard deviations of $C^N_{avg}$ at PBE/def2-SVP and the size of the molecules. The second column specifies the number of molecules within QM7 with the corresponding number of non-H atom.}
\label{tab:my-table}\\\hline
Non-H Atoms & Molecules & Mean of $C^N_{avg}$ & $\sigma(C^N_{avg})$ \\\hline
\endhead
1 & 1 & 9.168 & - \\
2 & 3 & 8.225 & 0.391 \\
3 & 12 & 7.93 & 0.551 \\
4 & 43 & 7.247 & 0.505 \\
5 & 158 & 6.766 & 0.466 \\
6 & 950 & 6.4 & 0.436 \\
7 & 5998 & 6.101 & 0.405 \\\hline
\end{longtable}

\bibliography{bibliography}